\documentclass[useAMS,usenatbib,rotating]{mn2e}
\usepackage{amssymb}     \usepackage{amsmath}    \usepackage{amsfonts}
\usepackage{xspace}    \usepackage{longtable}    \usepackage{rotating}
\usepackage{lscape}
\usepackage{times}

\title[Galaxy Zoo: Building low mass red sequence with PSGs]
  {Galaxy Zoo: Building the low-mass end of the red sequence with 
local post-starburst galaxies\thanks{This publication has been made 
possible by the participation of more than 250,000 volunteers in 
the Galaxy Zoo project. Their contributions are individually acknowledged 
at \texttt{http://www.galaxyzoo.org/Volunteers.aspx}.}}
\author[O.\ I.\ Wong et al.]
  {O.I.~Wong,$^{1,2}$ K.~Schawinski,$^{3,4,15}$ S.~Kaviraj,$^{5,6}$ 
   K.L.~Masters,$^{7,8}$  R.C.~Nichol,$^7$  
   \newauthor
   C.~Lintott,$^{9,6}$ W.C.~Keel,$^{10}$ D.~Darg,$^6$ S.P.~Bamford,$^{11}$  
   D.~Andreescu,$^{12}$  P.~Murray,$^{13}$
   \newauthor M.J.~Raddick,$^{14}$ A.~Szalay,$^{14}$ 
   D.~Thomas,$^{7}$ J.~VandenBerg$^{14}$\\
  $^1$CSIRO Astronomy \& Space Science, P.O. Box 76 Epping, NSW 1710, 
      Australia\\
  $^2$Astronomy Department, Yale University, P.O. Box 208101 New 
       Haven, CT 06520-8101, U.S.A.\\
  $^3$Department of Physics, Yale University, New Haven, CT 06511, U.S.A.\\
  $^4$Yale Center for Astronomy and Astrophysics, Yale University, 
     P.O. Box 208121, New Haven, CT 06520, U.S.A.\\
  $^5$Blackett Laboratory, Imperial College London, South Kensington Campus, 
     London SW7 2AZ, U.K.\\
  $^6$Oxford Astrophysics, Department of Physics, University of Oxford,
     Denys Wilkinson Building, Keble Road, Oxford, OX1 3RH, U.K.\\
  $^7$Institute for Cosmology and Gravitation, Dennis Sciama Building,
      University of Portsmouth, Burnaby Road, Portsmouth, PO1 3FX, U.K.\\
  $^8$South East Physics Network, SEPnet, (www.sepnet.ac.uk) \\   
  $^9$Adler Planetarium, 1300 S. Lakeshore Drive, Chicago, IL 60605, U.S.A.\\
  $^{10}$Department of Physics \& Astronomy, 206 Gallalee Hall, 514 University 
     Blvd., University of Alabama, Tuscaloosa, AL35487-0234, U.S.A. \\
  $^{11}$Centre for Astronomy \& Particle Theory, University of Nottingham, 
     University Park, Nottingham, NG7 2RD, U.K.\\
  $^{12}$LinkLab, 4506 Graystone Ave., Bronx, NY 10471, U.S.A.\\
  $^{13}$Fingerprint Digital Media, 9 Victoria Close, Newtownards, Co. Down, 
     Northern Ireland, BT23 7GY, U.K.\\
  $^{14}$Department of Physics and Astronomy, The Johns Hopkins University, 
     Homewood Campus, Baltimore, MD 21218, U.S.A.\\
  $^{15}$Einstein Fellow}
\date{Released 2011 Xxxxx XX}

\pagerange{\pageref{firstpage}--\pageref{lastpage}} \pubyear{2011}

\def\LaTeX{L\kern-.36em\raise.3ex\hbox{a}\kern-.15em
    T\kern-.1667em\lower.7ex\hbox{E}\kern-.125emX}

\begin{document}

\label{firstpage}

\maketitle

\begin{abstract}
We present a study of local post-starburst galaxies (PSGs) using the 
photometric and spectroscopic observations from the Sloan Digital Sky Survey 
(SDSS) and the results from the Galaxy Zoo project. We find that the majority
of our local PSG population have neither early- nor late-type morphologies but
 occupy a well-defined space within the colour--stellar mass diagram, most notably,
the low-mass end of the ``green valley'' below the transition mass thought
  to be the mass division between low-mass star-forming galaxies and high-mass 
passively-evolving bulge-dominated galaxies.   Our 
analysis  suggests that it is likely that a local PSG will quickly transform
 into ``red'', low-mass early-type galaxies as the stellar morphologies of the ``green'' 
PSGs  largely resemble that of the early-type galaxies within the same mass 
range.  We propose that the current population of PSGs represents  a 
population of galaxies which is  rapidly transitioning between the star-forming 
and the passively-evolving phases.  Subsequently, these PSGs will contribute
 towards the build-up of the low-mass end of the ``red sequence'' once the
current population of young stars fade and stars are no longer being formed.  
These results are consistent with the 
idea of ``downsizing'' where the build-up of smaller galaxies occurs at later 
epochs. 
\end{abstract}

\begin{keywords}
 galaxies: evolution
\end{keywords}

\section{Introduction}
Current research in galaxy evolution is still largely driven by our lack of  understanding of the link
between the two main types of galaxies observed in the sky. Many theories exist to explain the evolution 
between star-forming, gas-rich spiral galaxies (``late''-type) and non-star-forming, passively-evolving 
spheroid galaxies (``early''-type).  
Since this bimodal nature is highly correlated to the stellar age and star formation history of the 
individual galaxies, it is likely that we are observing two main stages of galaxy evolution   
\citep{strateva01,baldry04,baldry06,kauffmann04,drory09}; and that the transition between these two 
types occurs relatively quickly \citep{martin07}.  The colours and brightnesses (as defined by their
observed magnitudes) of all galaxies appear to be concentrated within two well-defined 
colour--magnitude regions.  The star-forming galaxies appear to populate a space dubbed the ``blue
cloud'' and the passively-evolving, non star-forming galaxies lie in a region called  the ``red sequence''.
Recent studies argue that local galaxies must migrate 
rapidly (within a Gyr) from the ``blue cloud'' to the ``red sequence''\footnote{There are a few cases 
where the evolution of individual galaxies moves from the ``red sequence'' to the ``blue cloud'' 
\citep[e.g.\ ][]{kannappan09,wei10}.  In these cases,  passively-evolving galaxies have accreted more gas 
 recently and are in the process of regrowing their stellar disk.  However, this ``red-to-blue'' mode of 
evolution is very unlikely to apply to our particular study because our PSGs are defined to  have no 
current on-going star formation.} due to the scarcity of galaxies within the intervening parameter space
 \citep[occasionally dubbed the ``green valley''; e.g.][]{schawinski07b}.  Therefore, valuable insights 
into galaxy evolution can be obtained by studying galaxies that appear to have intermediate 
properties and may be in the act of transitioning between the two main galaxy populations. 

Post-starburst galaxies (hereafter PSGs) or post-quenching galaxies \citep{yan09} such as 
``E+A'' or ``K+A'' galaxies, are galaxies which appear to have ceased current star formation, but 
still exhibit the spectral signature of recently-formed stars. In E+A galaxies, strong Balmer 
absorption lines are observed together with $\alpha$-element signatures such as  
Mg$_{5175}$, Fe$_{5270}$ and Ca$_{3934,3468}$ 
\citep{dressler83,dressler92,dressler99, dressler04,couch87,maclaren88,newberry90,fabricant91,abraham96,poggianti99,goto03,goto04,goto05,goto07}. Similar to E+A galaxies, K+A galaxies are PSGs which have a disk-like morphology
 \citep{couch94,dressler94,caldwell97,dressler99}.  
Although PSGs such as E+A galaxies are more common at higher redshifts \citep{wild09,tran04},
detailed high-resolution studies are only possible with a local population of PSGs since low redshift
 observations extend to lower surface brightness limits than that at earlier epochs.

Current studies favor the idea that PSGs are formed  via interactions or major mergers 
\citep{yamauchi08,blake04, bekki01} which trigger bursts of star formation. However, due to the effects
 of the merger interaction, the gas reservoir (from which stars are formed) is depleted and these PSGs 
eventually turn into bulge-dominated early-type galaxies once their star formation fades completely 
\citep[e.g.\ ][]{yang08}. \citet{kaviraj07} found that the quenching efficiency of star formation in
 less massive ($< 10^{10}$ M$_{\odot}$) and more massive  E+A galaxies ($> 10^{10}$M$_{\odot}$) is 
consistent with supernovae (SNe) and AGN being the main sources of negative feedback, respectively.

Using the Sloan Digital Sky Survey (SDSS), we assemble and analyze one of the largest, and 
most complete samples, of local PSGs to date.  This paper investigates the properties of the local PSGs
derived from the visual classifications of the Galaxy Zoo citizen science project \citep{lintott11,lintott08}.

Section 2 describes our sample selection and the sample properties are examined in Section 3. 
A discussion of our results and conclusions can be found in Section 4. The AB magnitude system is
used throughout this work.

\section{Our galaxy sample}

In this paper, we obtain the photometric and spectroscopic data from the Sloan Digital Sky Survey (SDSS) 
DR7 \citep{york00,abazajian09} for all objects classified as `galaxy' \citep{strauss02}. The main galaxy
emission line measurements are determined from the SDSS spectra using the {\tt{Gandalf}} IDL tool
by \citet{sarzi06}. To minimise the Malmquist bias, and create a volume and magnitude-limited (proxy for 
stellar-mass limited) sample of galaxies, we select all the galaxies within $0.02<z<0.05$ with $M_{z,Petro}<-19.5$ 
magnitudes.  We use the $z$-band since the reddest waveband provides the closest proxy to stellar mass.  
It should be noted that the results of this paper remain unchanged if the  $i$-band is used instead. 
As we aim to study the properties of all galaxies which have ceased star formation recently,
 we define a PSG to be a galaxy with a recently-truncated star formation history (i.e.\ where the 
observed H$\alpha$ emission line is weaker than four times the RMS level), while still exhibiting strong
 balmer absorption lines from recently-formed young stars (where the H$\delta$ equivalent width is wider
 than three Angstroms). Our strict H$\alpha$ criterion may result in the omission of a few PSG known to 
emit weak H$\alpha$ emission but the effects from the inclusion of a few H$\alpha$-emitting galaxies do not
change any of the results that we present in Section 3.

Studies such as \citet{balogh99,blake04,goto07}
 select for PSGs at higher redshifts ($0.5<z<1.0$) which exhibit very strong A-type stellar 
populations (where the observed H$\delta$ equivalent width is wider than five Angstroms). To include 
as many PSGs as possible into our sample, we imposed a more relaxed H$\delta$ 
equivalent width so we do not bias against galaxies with a low star formation rate (which results in 
weaker Balmer equivalent widths) prior to the cessation of star formation.

The [{\sc{Oii}}] forbidden lines were not used in the selection criteria because the H$\alpha$ emission 
line is a more accurate tracer of current star formation in the Local Universe.  The luminosity of the 
[{\sc{Oii}}] emission lines is not directly linked to the ionizing luminosity and the [{\sc{Oii}}] 
excitation is sensitive to the abundance and the ionisation state of the gas \citep{kennicutt98}.  
Moreover, the [{\sc{Oii}}]-derived star formation rates (SFRs) is sensitive to systematic errors from 
extinction and variations in the diffused gas fraction \citep{kennicutt98}. In starburst galaxies, the 
excitation of [{\sc{Oii}}] is much higher in the diffused ionised gas \citep{hunter90,hunter94,martin97}, 
and is able to double the $L$[{\sc{Oii}}]/SFR ratio in the integrated spectrum \citep{kennicutt92a}.

\begin{figure*}
\begin{center}
\begin{tabular}{ccc}
\includegraphics[scale=.4]{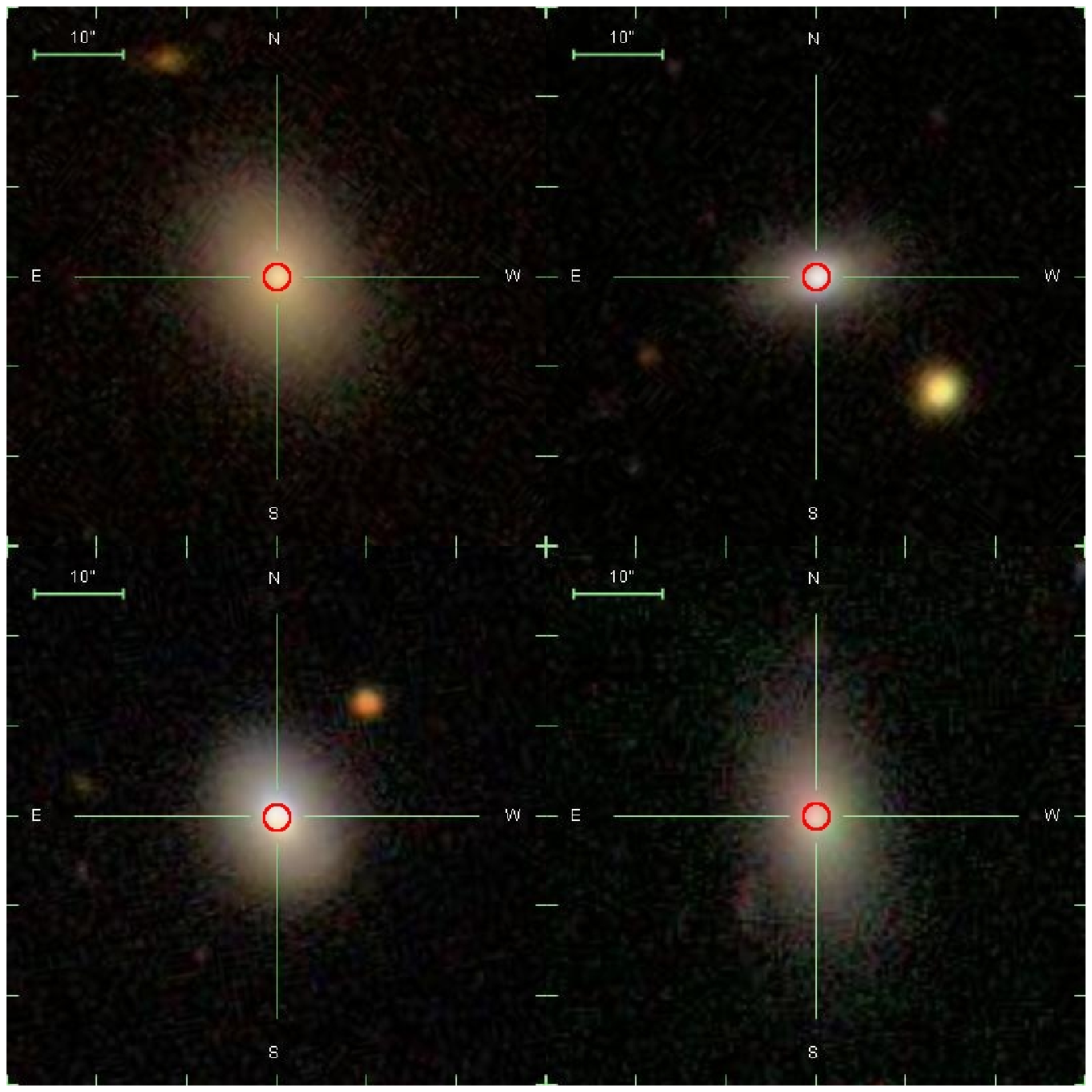} & \includegraphics[scale=.4]{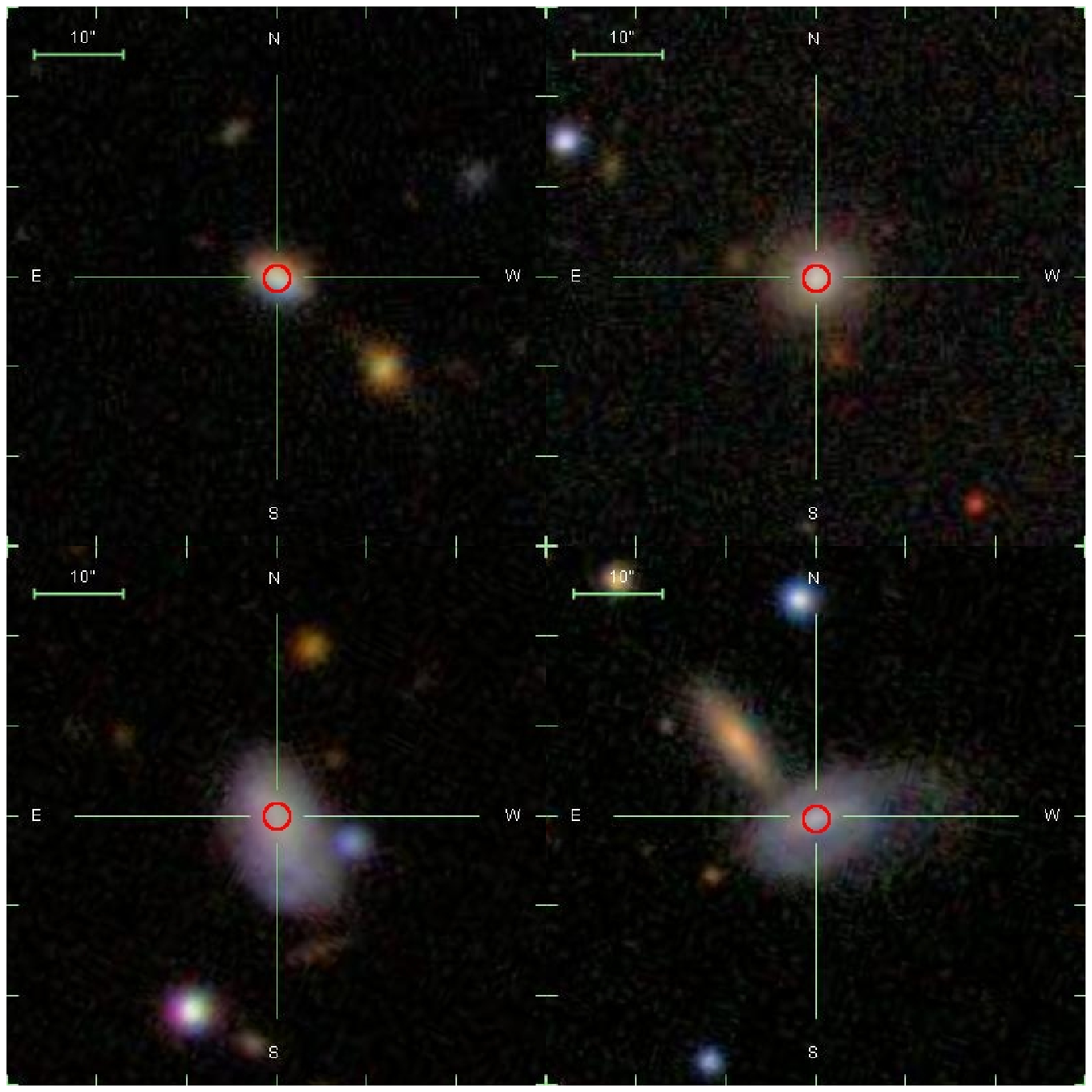}&\includegraphics[scale=.4]{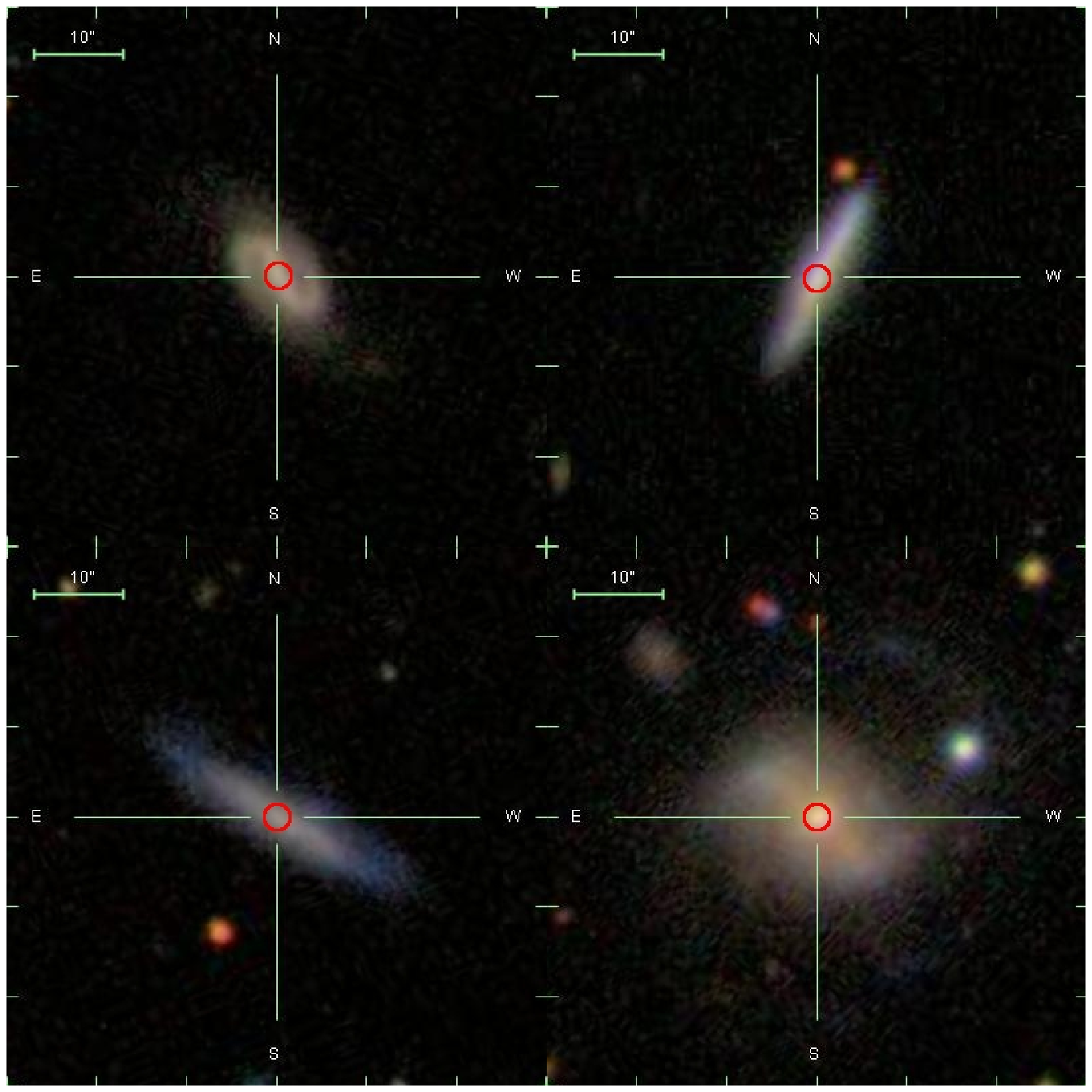}\\
\end{tabular}
\end{center}
\caption{Example SDSS colour images of 12 PSGs within our sample. Each image frame provides a  
48\arcsec$\times$48\arcsec\ field-of-view centered on the galaxy.  The left panel shows four examples of 
early-type PSGs, the middle panel shows four examples of intermediate-type PSGs and the right panel
shows four examples of late-type PSGs. The red circles show the SDSS fiber field-of-view where
each spectrum was obtained.}
\label{egsamp}
\end{figure*}

Of 47,573 galaxies within our selected volume, we find a total of 80 PSGs.  The general properties of our PSGs 
are listed in Table~\ref{genprops}. Figure~\ref{egsamp} shows the SDSS 
multicolour composite images for 12 random PSGs within our sample.  This figure exhibits the variety of 
morphologies and sizes within our PSG  sample.

Using the bootstrap resampling method to approximate the uncertainty, we find that our percentage of PSGs to the
total number of galaxies (within our specified volume out to $z\sim0.05$) is  0.17$^{+0.07}_{-0.05}$ \%. This
assumes that the uncertainty in our spectral line measurements is given by the signal-to-noise ratio (SNR).
  This PSG fraction is comparable to that of previous local PSG studies by \citet{goto05,goto07} and \citet{kaviraj07}. 
It should be noted that we do not find many of the E+A galaxies found by  \citet{goto05,goto07} (hereafter
 known as the ``G05'' sample) from the SDSS DR5 catalogue because we detect strong H$\alpha$ emission in these G05
 objects using the SDSS DR7 catalogue. Hence, these E+A galaxies appear to have current, on-going star formation 
and is inconsistent with our definition of a PSG.  Of the overlap galaxies between our full galaxy sample and that of 
the G05 sample, we find that our method of using the  Balmer absorption line strength finds every galaxy found via 
its absorption line equivalent width. This is consistent with the fact that there is a good correlation between the 
line strengths and the equivalent widths for galaxies of similar sizes.

\begin{table*}
\tiny{
\begin{center}
\caption{General properties of our PSG sample.}
\label{genprops}
\begin{tabular}{lcccccccccccl}
\hline 
\hline
SDSS object ID & RA  &Declination  &$m_{\rm{r}}$ & $u-r$ & $EW$(H$\delta$)& $S_{\rm{[\sc{OII}]}}$ & $S_{\rm{H}\alpha}$&log (M$_{\star}$) & $f_m$& $frac_{\rm{dev}}$  & Type & Comments\\ 
(1) & (2) & (3) & (4) & (5) & (6) & (7) & (8) & (9) & (10) &(11) &(12)\\
\hline 
  587731513679478898  &  01:14:47.2 & $+$00:37:55.7 &      16.75 &       2.00 &  3.4&$<7.9\times10^{-15}$&---&     9.95 &        0.000 &   0.595   &  I & ---      \\
  587731513681313986  &  01:31:37.9 & $+$00:48:51.7 &      16.52 &       1.37 &  5.0&$<2.4\times10^{-16}$&$<1.3\times10^{-17}$&     9.69 &        0.000 &   0.000   &  I & ---      \\
  587731511533961332  &  01:32:50.2 & $-$00:56:17.6 &      16.86 &       2.11 &  4.7&$<6.4\times10^{-15}$&$<1.5\times10^{-15}$&    10.03 &        0.000 &   1.000   &  I & blob      \\
  587727230522032233  &  01:37:16.4 & $-$09:29:06.5 &      16.63 &       2.25 &  3.7&$<1.2\times10^{-16}$&---&     9.96 &        0.000 &   0.546   &  I & blob       \\
  588015510358524013  &  02:20:38.7 & $+$00:54:09.1 &      15.15 &       0.97 &  4.1&$1.5\times10^{-13}$ &$<6.3\times10^{-15}$&     9.66 &        0.000 &   0.240   &  S & ---     \\
  587731512082956346  &  03:23:33.3 & $-$00:26:18.8 &      14.92 &       2.04 &  4.2&---&$<7.3\times10^{-14}$&     9.99 &        0.000 &   0.770   &  E & ---      \\
  587731514232996008  &  03:47:10.2 & $+$01:04:39.9 &      15.78 &       1.77 &  3.9&$<4.8\times10^{-16}$&$<4.7\times10^{-17}$&     9.97 &        0.000 &   0.007   &  I & ---      \\
  587732053779283987  &  08:43:20.7 & $+$37:13:27.6 &      16.24 &       2.26 &  3.7&$<5.5\times10^{-16}$&$<3.1\times10^{-16}$&    10.23 &        0.000 &   0.926   &  I & blob      \\
  587744874791370929  &  09:01:03.8 & $+$13:36:33.3 &      14.91 &       0.85 &  4.0&$1.7\times10^{-13}$ &$<1.2\times10^{-14}$&     9.82 &        0.188 &   0.505   &  I &  disturbed     \\
  587745403073855572  &  09:11:31.1 & $+$12:08:52.1 &      14.96 &       2.22 &  3.3&$<1.0\times10^{-14}$&$<2.4\times10^{-15}$&    10.43 &        0.047 &   1.000   &  I & disturbed      \\
  587745540514119842  &  09:25:03.2 & $+$13:03:58.0 &      16.56 &       2.06 &  3.0&$<3.5\times10^{-14}$&$<2.2\times10^{-15}$&     9.81 &        0.000 &   1.000   &  I &  blob     \\
  588016891707392070  &  09:29:34.6 & $+$33:47:51.1 &      15.28 &       1.22 &  4.3&$1.1\times10^{-13}$ &$<1.6\times10^{-14}$&     9.70 &        0.115 &   0.000   &  I & disturbed        \\
  587745243626405989  &  09:45:00.0 & $+$15:27:40.0 &      15.79 &       2.10 &  4.0&$<1.2\times10^{-14}$&$<4.2\times10^{-15}$&     10.25 &        0.000 &   0.609   &  I & blob      \\
  587735044693753946  &  09:46:29.9 & $+$39:02:19.7 &      16.72 &       1.34 &  3.9&$1.9\times10^{-13}$ &$<1.2\times10^{-14}$&     9.61 &        0.000 &   0.620   &  I & ---       \\
  587725074458804315  &  09:49:56.4 & $-$00:13:52.9 &      13.89 &       2.36 &  3.4&$<6.5\times10^{-16}$&$<2.7\times10^{-16}$&    11.35 &        0.095 &   1.000   &  I & disturbed    \\
  588848900973789221  &  10:04:29.8 & $+$00:41:20.2 &      16.49 &       2.06 &  3.2&$<1.8\times10^{-16}$&$<2.1\times10^{-16}$&     9.93 &        0.000 &   0.870   &  I& blob      \\
  587726032236183669  &  10:06:50.9 & $+$01:41:34.0 &      16.94 &       2.08 &  5.7&$<2.0\times10^{-16}$&---&     9.84 &        0.074 &   0.613   &  I & blob        \\
  587738948283334792  &  10:12:18.9 & $+$36:07:50.0 &      15.62 &       1.59 &  3.3&$1.1\times10^{-13}$ &$<1.9\times10^{-14}$&     9.96 &        0.000 &   0.964   &  I & blob      \\
  587741828579393617  &  10:21:25.9 & $+$21:32:45.8 &      16.00 &       1.90 &  4.1&---&$<6.2\times10^{-16}$&     9.91 &        0.000 &   0.002   &  S &   ---    \\
  587733080268931236  &  10:30:53.7 & $+$51:19:59.6 &      15.05 &       1.18 &  4.2&$4.9\times10^{-15}$&$<8.3\times10^{-16}$&    10.35 &        0.031 &   0.194   &  S & asymmetric  \\
  587728947978436717  &  10:42:32.3 & $-$00:41:58.3 &      15.83 &       2.07 &  4.3&$<2.9\times10^{-16}$&$<7.5\times10^{-16}$&    10.10 &        0.000 &   0.994   &  I & blob        \\
  587729386611212446  &  10:53:05.4 & $+$57:51:54.2 &      15.60 &       1.10 &  3.8&$3.7\times10^{-16}$&$<1.5\times10^{-16}$&    10.05 &        0.000 &   0.054   &  I & ---        \\
  587734894357381314  &  11:00:48.5 & $+$10:31:19.0 &      16.44 &       2.23 &  4.2&$<6.1\times10^{-15}$&$<1.6\times10^{-15}$&    10.12 &        0.000 &   0.864   &  I &asymmetric      \\
  587741489834754106  &  11:10:33.9 & $+$28:29:33.3 &      15.73 &       2.07 &  4.1&---&$<1.4\times10^{-14}$&     10.00 &        0.000 &   0.985   &  I &  blob     \\
  588848898833842380  &  11:13:28.0 & $-$00:54:09.5 &      16.62 &       2.03 &  4.9&$<2.5\times10^{-14}$&$<3.7\times10^{-15}$&     9.57 &        0.000 &   0.771   &  I & blob      \\
  587732580982521898  &  11:19:07.6 & $+$58:03:14.3 &      14.17 &       2.16 &  4.7&$1.6\times10^{-13}$&$<2.9\times10^{-14}$&    10.74 &        0.000 &   1.000   &  I & asymmetric      \\
  587739405703577638  &  11:26:53.7 & $+$33:07:09.5 &      14.86 &       1.21 &  4.5&$1.3\times10^{-13}$ &$<7.3\times10^{-15}$&     9.92 &        0.108 &   0.190   &  S & --- \\
  587732482746548338  &  11:35:32.0 & $+$48:56:38.5 &      15.99 &       2.19 &  4.0&$<6.5\times10^{-15}$&$<1.3\times10^{-15}$&     9.76 &        0.017 &   0.866   &  I & disturbed      \\
  587741726574444657  &  11:36:55.2 & $+$24:53:25.5 &      14.92 &       1.73 &  5.5&$1.4\times10^{-13}$ &$ <5.0\times10^{-11}$  &   9.98 &        0.027 &   0.200   &  I &  disturbed     \\
  587742573224657026  &  11:43:47.8 & $+$20:21:48.0 &      15.61 &       1.83 &  6.7&$<8.2\times10^{-15}$&---&     9.56 &        0.000 &   0.305   &  S & disturbed   \\
  588017111833182222  &  12:21:05.7 & $+$47:58:51.9 &      16.54 &       2.18 &  3.6&$<7.2\times10^{-14}$&$<2.2\times10^{-15}$&    10.02 &        0.000 &   0.978   &  I &  ---     \\
  588017730836561977  &  12:26:41.6 & $+$08:44:32.2 &      15.42 &       2.45 &  6.5&$<8.9\times10^{-13}$ &$<4.9\times10^{-14}$&    10.38 &        0.000 &   0.906   &  I &  ---     \\
  587726033325850747  &  12:32:18.9 & $+$03:00:09.8 &      16.82 &       2.00 &  3.3&$<6.2\times10^{-14}$&---&    10.10 &        0.000 &   1.000   &  I & blob        \\
  587725816414142618  &  12:32:23.3 & $+$66:25:36.6 &      16.90 &       2.26 &  3.1&$<1.0\times10^{-16}$&---&     9.82 &        0.032 &   0.528   &  I & blob     \\
  587734892220252284  &  12:37:18.0 & $+$09:32:09.0 &      14.55 &       1.89 &  3.0&$<3.4\times10^{-16}$&---&     9.84 &        0.000 &   0.979   &  E & ---      \\
  587739096454332452  &  12:38:52.8 & $+$36:32:05.7 &      15.74 &       2.15 &  3.7&$<1.2\times10^{-16}$&$<1.3\times10^{-17}$&    10.19 &        0.000 &   1.000   &  E & ---     \\
  587741602572599308  &  12:57:17.8 & $+$27:48:39.3 &      15.40 &       1.97 &  3.9&$<1.7\times10^{-15}$&$<4.4\times10^{-18}$&     9.74 &        0.000 &   0.211   &  I &  disturbed     \\
  587741722823557134  &  12:57:21.7 & $+$27:52:49.5 &      15.23 &       2.10 &  3.6&$<5.1\times10^{-15}$&$<1.9\times10^{-17}$&     9.94 &        0.000 &   0.873   &  E &  ---   \\
  587741722286686489  &  12:57:45.7 & $+$27:25:45.6 &      15.92 &       1.94 &  3.9&$<1.3\times10^{-13}$&$<1.1\times10^{-14}$&     9.56 &        0.000 &   0.254   &  S &  ---     \\
  587741721749881039  &  12:58:12.0 & $+$27:07:39.5 &      16.60 &       2.31 &  3.5&$<3.0\times10^{-14}$&---&    10.06 &        0.000 &   0.806   &  S & ---      \\
  587741722823754043  &  12:59:39.5 & $+$27:51:16.6 &      16.57 &       2.20 &  4.2&$<3.7\times10^{-17}$&$<1.1\times10^{-17}$&     9.95 &        0.454 &   0.343   &  I &  disturbed     \\
  587741722823819345  &  13:00:10.2 & $+$27:51:50.2 &      15.85 &       2.22 &  3.4&$<2.8\times10^{-15}$&---&    10.00 &        0.097 &   0.357   &  I &  disturbed    \\
  587741722286948519  &  13:00:29.2 & $+$27:30:53.4 &      15.53 &       1.63 &  8.5&$<1.1\times10^{-15}$&$<9.6\times10^{-17}$&     9.55 &        0.125 &   0.224   &  I &  disturbed     \\
  587739719754514434  &  13:04:22.7 & $+$28:48:38.8 &      14.33 &       0.86 &  4.4&$<2.9\times10^{-12}$&$<4.2\times10^{-14}$&    10.15 &        0.365 &   0.643   &  S & disturbed \\
   587733195160485975  &  13:05:25.8 & $+$53:35:30.3 &      14.29 &       2.15 & 4.6&$<4.8\times10^{-16}$&$<1.4\times10^{-16}$&     10.75 &        0.000 &   1.000   &  E & ---      \\
  587722982822379684  &  13:17:59.6 & $-$00:17:43.2 &      13.29 &       1.22 &  5.2&$<2.1\times10^{-14}$ & $<5.1\times10^{-14}$&   10.44 &        0.139 &   0.477   &  S & asymmetric  \\
  587729773681377291  &  13:37:44.1 & $-$02:10:27.9 &      14.61 &       0.72 &  3.5&---&$<2.5\times10^{-14}$&   10.41 &        0.000 &   0.338   &  S & ---      \\
   587726032797565040  &  13:50:51.0 & $+$02:19:38.5 &      15.62 &       2.11 & 4.7&$<3.2\times10^{-16}$&$<8.7\times10^{-17}$&      9.97 &        0.167 &   1.000   &  I & disturbed        \\
  587730021178867826  &  14:18:51.0 & $+$05:28:14.2 &      15.92 &       2.09 &  3.0&$<7.3\times10^{-15}$&$<2.6\times10^{-15}$&    9.90 &        0.000 &   0.961   &  I &  blob     \\
  587739827130007710  &  14:39:08.5 & $+$22:17:42.4 &      15.42 &       0.84 &  4.5&$2.0\times10^{-14}$&$<1.1\times10^{-15}$&    9.75 &        0.136 &   0.000   &  I & asymmetric      \\
  587739408405626997  &  14:39:30.4 & $+$30:52:49.1 &      16.14 &       1.97 &  4.2&$<5.0\times10^{-14}$&$<1.3\times10^{-15}$&     9.97 &        0.153 &   1.000   &  I & disturbed     \\
  587739379916210254  &  14:54:24.7 & $+$27:39:38.7 &      16.23 &       2.04 &  3.9&$<1.8\times10^{-16}$&$<4.2\times10^{-17}$&     9.98 &        0.000 &   1.000   &  I & blob      \\
 587726032268362094  &  15:01:24.0 & $+$01:39:25.0 &      15.90 &       1.92 &   3.9&---&$<4.5\times10^{-15}$&    9.95 &        0.000 &   0.682   &  I & blob        \\
  588017949366157511  &  15:03:39.4 & $+$35:27:30.5 &      16.14 &       0.94 &  3.9&$1.8\times10^{-13}$ &$<6.4\times10^{-15}$&     9.60 &        0.000 &   0.000   &  I & ---      \\
  587736975271067720  &  15:12:04.1 & $+$28:25:14.1 &      14.26 &       1.20 &  4.2&$1.3\times10^{-15}$&$<4.7\times10^{-15}$&    10.43 &        0.015 &   0.772   &  I & asymmetric      \\
  588017991773978739  &  15:14:29.9 & $+$07:35:46.8 &      15.45 &       1.97 &  4.4&$<4.7\times10^{-14}$&$<5.5\times10^{-15}$&    10.27 &        0.000 &   0.918   &  I & blob      \\
  587736543098568963  &  15:21:08.6 & $+$07:37:53.9 &      16.40 &       2.05 &  4.9&$<1.7\times10^{-14}$&$<2.5\times10^{-15}$&    10.12 &        0.000 &   0.920   &  I & blob      \\
  587742575924805956  &  15:23:22.8 & $+$13:26:19.2 &      15.91 &       1.38 &  3.4&$2.8\times10^{-14}$ &$<7.7\times10^{-15}$&     9.99 &        0.000 &   0.038   &  I &  disturbed     \\
  587736477596975155  &  15:23:43.0 & $+$08:27:29.0 &      15.86 &       2.06 &  3.0&$<5.2\times10^{-16}$&---&     9.81 &        0.000 &   1.000   &  I & ---      \\
  587739810494218509  &  15:24:25.6 & $+$19:42:59.5 &      14.96 &       1.13 &  5.4&$1.2\times10^{-14}$&$<1.4\times10^{-15}$&     9.73 &        0.027 &   0.031   &  S &  L; disturbed   \\
  587733603187556413  &  15:26:21.9 & $+$48:42:14.4 &      15.88 &       1.58 &  3.9&$1.4\times10^{-13}$ &$<5.7\times10^{-14}$&    10.07 &        0.310 &   1.000   &  I & ---      \\
  588017703489634681  &  15:26:53.3 & $+$08:32:07.7 &      16.64 &       2.21 &  4.7&$<1.3\times10^{-13}$&$<5.7\times10^{-15}$&    10.00 &        0.000 &   0.359   &  I &  blob     \\
  587730021724062009  &  15:34:53.4 & $+$04:16:60.0 &      15.66 &       1.43 &  3.5&$1.9\times10^{-15}$&$<7.4\times10^{-16}$&    10.00 &        0.000 &   0.181   &  I &  disturbed      \\
  587739721917595889  &  15:39:19.9 & $+$21:21:30.0 &      16.76 &       2.05 &  3.1&$<3.9\times10^{-15}$&$<8.8\times10^{-16}$&     9.86 &        0.000 &   1.000   &  I &   blob    \\
  588017704565276972  &  15:44:31.7 & $+$08:35:07.4 &      16.52 &       1.93 &  3.6&$<6.4\times10^{-14}$&$<3.3\times10^{-15}$&     9.75 &        0.000 &   0.717   &  S &  ---     \\
 588011101570662666  &  16:00:17.3 & $+$46:51:35.6 &      16.57 &       2.28 &   3.3&$<9.8\times10^{-14}$ &$<8.1\times10^{-15}$&   10.29 &        0.000 &   0.779   &  E & blob      \\
  587739828212793702  &  16:03:15.7 & $+$16:19:07.9 &      15.86 &       2.09 &  3.6&$3.1\times10^{-14}$&$<4.5\times10^{-15}$&    10.07 &        0.044 &   0.814   &  I & disturbed      \\
  587739809961804053  &  16:03:25.9 & $+$15:38:48.2 &      15.43 &       0.84 &  4.0&$5.7\times10^{-14}$ &$<4.2\times10^{-15}$&     9.55 &        0.000 &   0.106   &  S &  L   \\
   587729227152752660  &  16:03:44.5 & $+$52:24:12.6 &      15.57 &       2.23 & 4.3&$<5.8\times10^{-15}$&$<1.2\times10^{-15}$&     10.42 &        0.000 &   1.000   & E &  ---     \\
  587739707420901722  &  16:04:38.6 & $+$17:36:35.0 &      16.61 &       2.00 &  3.8&$<3.3\times10^{-17}$&$<1.5\times10^{-16}$&     9.46 &        0.186 &   0.499   &  I & disturbed      \\
  587725817501712592  &  16:16:52.9 & $+$50:57:04.6 &      14.69 &       2.73 &  4.0&---&$<6.7\times10^{-15}$&    11.47 &        0.000 &   0.734   &  E & ---     \\
  587735666928844900  &  16:25:18.3 & $+$37:56:40.6 &      15.03 &       2.12 &  3.2&$<1.0\times10^{-15}$&$<1.6\times10^{-16}$&    10.15 &        0.000 &   1.000   &  I & ---      \\
  588018090548920817  &  16:41:33.5 & $+$24:42:14.2 &      16.01 &       2.02 &  4.1&$<6.9\times10^{-15}$ &$<2.3\times10^{-15}$&     10.04 &        0.000 &   0.849   &  I & blob      \\
  588018253228408956  &  16:53:14.4 & $+$25:07:59.2 &      14.01 &       1.45 &  4.3&$1.1\times10^{-13}$ & $<9.3\times10^{-14}$&    10.44 &        0.000 &   0.692   &  S &  disturbed     \\
  588018254302609790  &  16:59:33.1 & $+$24:55:59.9 &      16.17 &       2.36 &  3.0&$<7.3\times10^{-16}$&$<2.2\times10^{-16}$&    10.28 &        0.000 &   0.717   &  I &  blob     \\
  587733432998559923  &  17:04:51.3 & $+$25:03:18.4 &      16.51 &       2.51 &  8.3&$<1.1\times10^{-14}$&$<2.4\times10^{-15}$&    10.45 &        0.068 &   1.000   &  I &  blob     \\
  587725503947735340  &  17:14:07.2 & $+$57:28:38.7 &      16.00 &       2.10 &  3.2&$<1.4\times10^{-14}$&$<8.2\times10^{-16}$&     9.83 &        0.000 &   0.129   &  S & ---   \\
  587731174382502294  &  21:19:48.2 & $+$00:40:21.8 &      16.37 &       1.83 &  5.5&$<3.7\times10^{-16}$&$<5.5\times10^{-17}$&     9.91 &        0.000 &   0.880   &  I & ---        \\
  587727213347799419  &  21:33:56.5 & $-$07:54:08.7 &      15.22 &       1.14 &  4.4&$<2.7\times10^{-16}$&$<5.5\times10^{-17}$&     9.75 &        0.000 &   0.000   &  I & ---       \\
  587731185117954089  &  22:08:06.1 & $-$00:54:25.0 &      15.10 &       2.06 &  4.3&$<4.9\times10^{-16}$&$<3.3\times10^{-16}$&    10.54 &        0.036 &   1.000   &  E & disturbed      \\
\hline  
\hline
\end{tabular}
\end{center}
\begin{flushleft}
Col.\ (1): SDSS object identification.  Col.\ (2): Galaxy center's right ascension.  Col.\ (3): Galaxy center's declination.
Col.\ (4):  $r$-band magnitude.  Col.\ (5): $u-r$ colour. Col.\ (6): H$\delta$ equivalent width (Angstroms). 
Col.\ (7): Measured {\sc[Oii]} flux densities in units of $10^{-17}$ ergs s$^{-1}$ cm$^{-2}$ \AA$^{-1}$.  The  upper limits are defined to be four times the RMS noise.  Col.\ (8): Estimated upper limits for the H$\alpha$ flux densities in 
units of $10^{-17}$ ergs s$^{-1}$ cm$^{-2}$ \AA$^{-1}$ as defined to be four times the RMS noise. 
Col.\ (9): Log of the stellar mass obtained from \citet{schawinski07b} which were determined by fitting the five SDSS 
photometric bands to model star formation histories  from \citet{maraston98, maraston05} stellar models . Col.\ (10): Weighted 
merger parameter \citep{darg10}. Col.\ (11): The SDSS parameter  which gives the fraction of light (in the $r$-band) fitted by 
a de Vaucouleurs profile. Col.\ (12): Description of galaxy type where `S' represents late-type spiral galaxies, `E' represents 
early-type galaxies and `I' represent intermediate-type galaxies.  Col.\ (10): Comments from visual inspection of each galaxy.  
A `disturbed' comment indicates that the stellar morphology appears to have been disrupted by tidal interaction with features such 
as stellar tails/rings or with neighboring galaxies at close proximities. An `asymmetric' defines an observed asymmetric stellar disk.  
A `---' comment represents galaxies whose morphologies appear fairly regular, while a `blob' comment refers to galaxy morphologies 
which are small, compact and fairly spherical in appearance. An 'L' indicates that this galaxy exhibits spectral signatures of a LINER. 
\end{flushleft}}
\end{table*}

\section{Results}

\subsection{Morphological properties}

\subsubsection{Morphological classification using Galaxy Zoo}
Galaxy morphologies are derived from the Galaxy Zoo 1 project  \citep{lintott11}.  From the multiple independent
 inspections (and classifications) made for each galaxy, the accuracy of the classifications for individual galaxies 
can be determined by imposing a required level of agreement among the classifiers.  Following the definition 
of the {\tt{clean}} sample from \citet{land08}, we require a minimum of an 80\% majority agreement on the 
morphology for each object.  Spiral galaxies (from an Sa to an Sd morphology) are classified as {\em{late-type}} 
objects, while all spheroids  (including lenticular  galaxies) are classed as {\em{early-type}}.    Galaxy 
 morphologies which are neither ``early''- nor ``late'' are classed as {\em{intermediate-type}} which include 
galaxies with irregular/disturbed and merging morphologies. 

\begin{table}
\begin{center}
\caption{Distributions of galaxy morphologies and star formation properties of our entire
local volume sample and that of the PSG sample. The distributions are listed as percentages 
and the number of galaxies are given in parentheses. }
\label{morph1}
\scriptsize{
\begin{tabular}{lcccc}
\hline
\hline
\multicolumn{4}{c}{Entire Galaxy Sample} & PSG sample \\
 & Non-SF & SF & Total & \\
\hline
Early-type &     50\% (5148)   &   10\% (3770)     &     19\% (8918)   & 16\% (13) \\
Intermediate &   46\% (4841)   &   47\% (17589)    &             47\% (22430)   & 74\% (59)\\
Late-type  &   4\% (388)    &   43\% (15837)     &              34\% (16225)    & 10\% (8) \\
Total &     100\% (10377)      &   100\% (37196)     &               100\% (47573)  & 100\% (80)\\
\hline
\hline
\end{tabular}}
\end{center}
\end{table}

Relative to the entire local volume sample, there is a smaller fraction of PSGs with spiral or late-type
morphologies. We find that 74 ($\pm10)$\% of our PSG sample appear to have intermediate morphologies, while 
16 ($\pm 5$)\%  and 10 ($\pm 4$)\% of our PSGs are classed as early- and late-types, respectively. It should
 be noted that a colour bias in morphology votes (by the Galaxy Zoo citizen scientists) is unlikely to occur
 since our PSG sample is dominated by intermediate-type morphologies.  In addition, investigations by
\citet{lintott08} found colour bias to not be a significant effect on the final morphological classifications.

Similar to the results of \citet{baldry04,driver06,bamford09}, we find that the majority of the spheroidal/elliptical
galaxies within our local volume sample exhibit non-star-forming (non-SF) properties, while, the star-forming
(SF) galaxies tended to have spiral morphologies. In this paper, we classify a galaxy as star-forming if its
 H$\alpha$ emission line is stronger than four times the RMS level.  As expected from our PSG selection criteria 
of galaxies with recently-truncated star formation, we find that the PSGs appear to have a morphology distribution 
which appears intermediate to that of SF and non-SF galaxies within the control sample.
Table~\ref{morph1} lists the distribution of morphologies and star-forming properties  of both the local volume
control sample and the PSG sample.

In addition to the Galaxy Zoo classifications, we visually inspected each PSG to confirm the morphologies.  
Consistent with the fact that a majority of the PSGs have intermediate-type morphologies, we find disturbed galaxy 
morphologies which resemble neither early- nor late-type galaxies.      Depending on the size 
of the galaxy, the 3\arcsec\ fibers (from which the SDSS spectra are obtained) may only correspond to the central 
1.2--2.9 kpc of galaxies within our redshift range. In our sample, the observed spectral properties may not be 
representative for approximately 5\% of our sample where the outer 
galaxy regions are much  greater than 3\arcsec\ and appear to be bluer than the central region.

\begin{figure}
\begin{flushleft}
\includegraphics[scale=0.43]{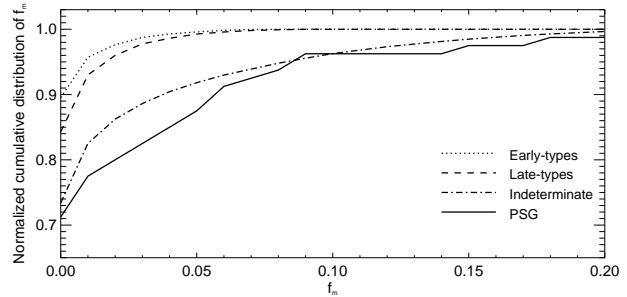}
\end{flushleft}
\caption{Peak-normalized cumulative distributions of $f_{\rm{m}}$ for our PSG sample (black solid line) and
the different types of galaxies within the control sample. }
\label{fmdist} 
\end{figure}

The $f_{\rm{m}}$ parameter is a quantification of the merger properties of Galaxy Zoo 1 and is defined to be the
 weighted-merger-vote fraction where the fraction of merger votes is multiplied by a weighting factor 
\citep[$W$; ][]{lintott08} to account for the quality of the particular voters who have assessed each galaxy 
\citep{darg10}. Assuming that $f_{\rm{m}} > 0.4$ describes a merger \citep{darg10}, we do not find many mergers within our PSG sample even though a large number of the PSGs 
appear asymmetrical or disturbed.  A comparison of the distribution of $f_{\rm{m}}$ parameters of our PSG sample
 to those of the early- and late-types of the entire galaxy sample yields Kolmogorov-Smirnov (KS) probabilities
which indicate that there are no significant differences between any of the $f_{\rm{m}}$ distributions. Figure~\ref{fmdist} 
shows the normalized cumulative distributions of $f_{\rm{m}}$ for the different galaxy types.


\subsubsection{Concentration}

As the majority of our PSG sample consists of intermediate-type morphologies, further investigation of the stellar
structure will reveal quantitatively whether these PSGs truly have intermediate-type morphologies (possibly due to past interactions) or are 
 similar in structure to the early- or late-type galaxies within the same volume.  We measure the  concentration 
index  using the ratio between the $R90$ and $R50$ parameters which are the radii where 90\% and 
50\% of the total Petrosian\footnote{A Petrosian radius is where the mean local surface brightness (within the local 
annulus) is equal to a constant fraction of the mean surface brightness within that radius \citep{strauss02}.} flux have
been measured in the $i$-band.  Therefore bulge-dominated galaxies will have greater $R90/R50$  ratios, while
disk-dominated galaxies will have smaller values.

To determine the end-products of our PSG sample, we compare the peak-normalised distributions of concentration
 indices for our population of  PSGs  to the early- and late-type population of galaxies with 
log (M$_{\star}$) $<10.5$ M$_{\odot}$.  In agreement with \citet{strateva01}, we find that the division between
early- and late-type galaxies is where the $R90$ is approximately 2.6 times greater than the $R50$.
The distribution of concentration indices for our PSG sample is very similar  to that of the early-type galaxies.

\begin{figure}
\begin{center}
\includegraphics[scale=0.45]{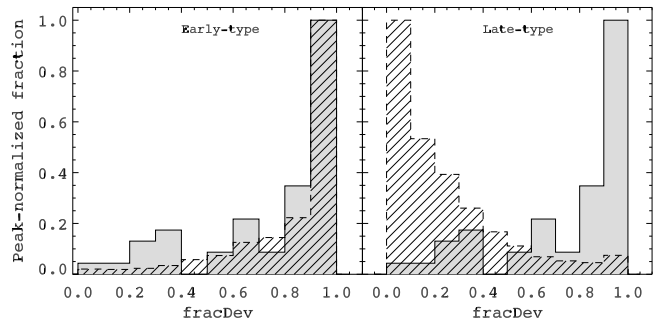}
\end{center}
\caption{Peak-normalised distributions of {\tt{fracDev}} (which describes the fraction of light from a fit to a
de Vaucouleurs profile). The gray shaded histograms show the {\tt{fracDev}} distribution for PSGs.  The 
striped histogram show the {\tt{fracDev}} distribution for the early-type galaxies (in the left panel) and
the distribution for the late-type galaxies (in the right panel).}
\label{fracdevfig}
\end{figure}

On the other hand, \citet{masters10} showed that early-type galaxies classified in such a manner may be 
contaminated by up to 50\% by edge-on spirals and that the SDSS {\tt{fracDev}} parameter ({\tt{fracDev}}; which 
gives the fraction of light  fitted by a de Vaucouleurs profile) provides a better differentiation between the early- 
and late-type galaxies as typical early-type elliptical galaxies are traditionally characterised by a de Vaucouleurs 
profile and as such have an {\tt{fracDev}} $ > 0.5$.  As such, we show the distribution of {\tt{fracDev}} for our 
``green'' PSGs and low-mass early- and late-type sample in Figure~\ref{fracdevfig}.  The PSG distribution is shown
by the gray-shaded histogram, while the striped histogram in the left-panel represents the distribution for the
early-types and the distribution for late-types are represented by the striped histogram in the right panel.
A quantitative comparison using the KS test yields a KS probability of 0.68 that the {\tt{fracDev}} distributions of 
PSGs and early-types are derived from the same parent sample, while  the {\tt{fracDev}} distributions of PSGs and 
late-types yield a KS probability of $<$0.001.  

Therefore, we find that the structural stellar morphologies of the PSGs within the ``green valley'' appear to be
more closely related to the morphologies of low-mass early-type galaxies even though star formation has only been
truncated recently.

\subsection{Colour and stellar mass}

We derive the $u-r$ colours of our sample using the {\tt{modelMags}} \citep[from SDSS DR7; ][]{abazajian09} 
which are determined from the best fit of each galaxy profile to the linear combination of the exponential and the 
de Vaucouleurs profiles.
In addition, these magnitude measurements are corrected for dust attenuation using the models of \citet{calzetti00}.
The ``green valley'' of our sample's colour distribution is defined to be within the nominal colour range of $1.8<u-r<2.3$.
The stellar mass estimates for each galaxy are measured by fitting the five optical wavebands from SDSS to star formation
history libraries generated from stellar models of \citet{maraston98,maraston05}. The uncertainties in our derived
stellar masses are dominated by the inherent uncertainties within the stellar models used. More details on the 
parameterization of the star formation histories  and the fitting process can be found in \citet{schawinski10a} and 
references within.

\begin{figure*}
\raggedright{
\includegraphics[angle=90,scale=0.535]{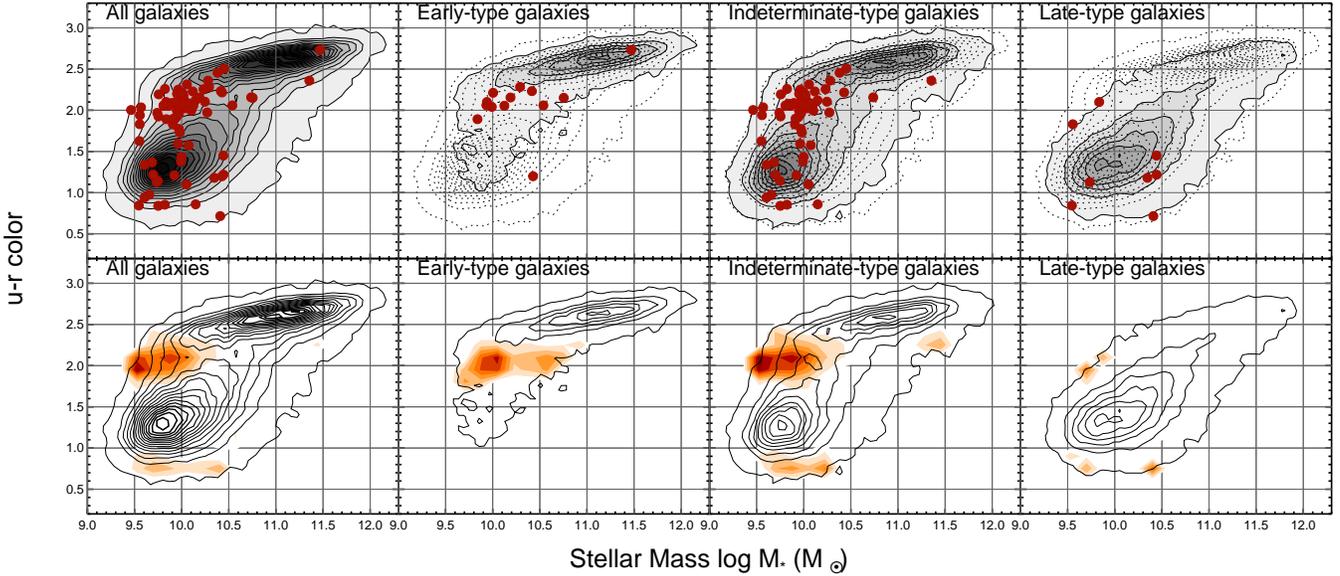}
\caption{The $u-r$ colour versus stellar mass distribution of our the SDSS galaxy sample are shown by the contours.  
The top row shows PSG sample (solid red circles) overplotted onto the distribution of SDSS galaxies of a particular
type (demarcated by the solid line contours). The panels from left to right show the distributions: for all galaxy types, 
for early-type galaxies, for intermediate-type galaxies and for late-type galaxies.
Within the top row, the distribution of SDSS galaxies for all galaxy types
are also shown as a comparison by dotted-lined contours within the early-, intermediate and late-type galaxy panels.
The bottom row shows the number fraction of our PSG sample to the galaxy sample of a particular type in a given 
colour--mass bin as a solid orange contour map overlaid onto the contours marking the distribution of SDSS galaxies. 
The maximum number fractions for all galaxy types, early-types, intermediate-types, late-types are 3.0\%, 2.6\%, 
3.0\% and 1.9\%; respectively.  }
\label{colourmass}
}
\end{figure*}

The top row of panels in Figure ~\ref{colourmass} shows the colour--stellar mass distribution of our PSG sample as red 
solid circles and the SDSS galaxy sample of a particular type is represented by the black solid contours. Black 
dotted-lined contours represent the distribution of the entire sample regardless of galaxy-type and are shown for 
comparison purposes.  The bottom row of panels in Figure~\ref{colourmass} shows the distribution of the fraction of PSGs 
to the number of galaxies (of a particular type) in a given colour--stellar mass bin as a solid orange-shaded 
contour map overlaid on the solid contours showing the distribution of the SDSS galaxies of  a particular type.  

\begin{figure}
\begin{center}
\includegraphics[scale=0.5]{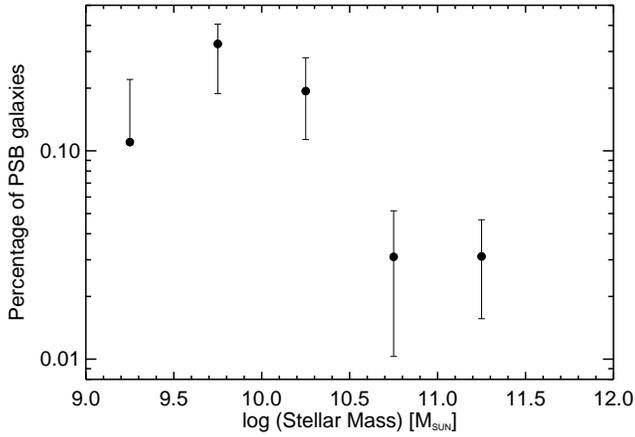}
\end{center}
\caption{Percentage of PSGs as a function of stellar mass. The uncertainties are given by the 10$^{\rm{th}}$ and 90$^{\rm{th}}$ percentile
values derived from bootstrap resampling of our data. }
\label{fracerrsmassbin}
\end{figure}

As can be seen from the top row panels of Figure~\ref{colourmass}, our PSG sample appears to be spread over a fairly 
large $u-r$ colour range even though a majority (94\%) of the PSGs have stellar masses below the transition mass of 
log $M_{\star} <$ 10.5 M$_{\odot}$ which separate the low-mass star-forming galaxies from the high-mass passively-evolving 
bulge-dominated galaxies \citep{kauffmann03}. In fact, we do not find any PSG with a log $M_{\star} >$  11.5 M$_{\odot}$. 
Such a stellar mass limit in our PSG sample is more clearly illustrated by the number fractions of PSGs to 
the number of galaxies within a particular colour--stellar mass range (see the bottom panels of Figure~\ref{colourmass}).
  Figure~\ref{fracerrsmassbin} shows the percentage of PSGs per 10$^{0.5}$ M$_{\odot}$  bins. The average 
percentage of PSGs in the log $M_{\star}$ bins between  9.5 and 10.5 M$_{\odot}$  is $\sim8$ times greater than that 
between the log $M_{\star}$ bins between 10.5 and 11.5 M$_{\odot}$ (with a 3$\sigma$ significance). The uncertainties in the stellar
masses are dominated by the uncertainties in the stellar population models and can be up to 10$^{0.3}$ M$_{\odot}$  in
stellar masses \citep[e.g.\ ][]{conroy09}.

One possible reason for the lack of high mass PSGs is likely to be because our sample is restricted to a very
 local volume. As such, the probability of finding massive galaxies in such a local volume is much less than at 
higher redshifts. Galaxies with log stellar masses greater than 11.5  account for only 3.8\% of the entire galaxy 
sample. In addition, high-redshift surveys of E+A galaxies are not as sensitive to smaller (and fainter) galaxies.  
As such, the E+A galaxies found at $z\sim0.1$ are most likely the biggest and brightest types of post-starburst 
galaxies. For example, the PSG sample at $z=0.8$ in \citet{yan09} have stellar masses greater than 
$10^{10.6}$ M$_{\odot}$. In addition, low redshift PSGs are also more closely associated to less massive blue 
galaxies than high redshift PSGs which are more similar to massive red galaxies \citep[e.g.\ ][]{yan09}. Therefore,  
it is not surprising that our local sample of PSGs is dominated by low mass objects.

In addition to the apparent stellar mass limit of  our PSG population, the bottom panels of  Figure~\ref{colourmass} 
also show that a significant fraction (61\%) of our PSG sample reside within the ``green valley'', while 31\% 
and 8\% reside in the ``blue cloud'' ($u-r < 1.8$) and the ``red sequence'', respectively.  Similarly, we find that 
the percentage of  PSGs within the ``green valley'' (0.6\%) is 17 times and 6 times greater than the fraction 
of PSGs within the ``red sequence'' and the ``blue cloud'', respectively. Hence, we propose that local PSGs occupy 
a well-defined position in the low-mass end of the ``green valley''.   Consistent with the idea of galaxy formation
 downsizing, we postulate that local PSGs  will transform into passively-evolving ``red'' galaxies
 and contribute towards the build-up of  the low-mass end of the ``red sequence'' if star formation has indeed ceased.


\subsection{Environment}
Using the adaptive Gaussian environment parameter, $\rho_{\rm{g}}$ \citep[which provides a measure of 
the number and proximity of galaxies around a point in space, see][]{schawinski07a},  we study the 
environment properties of our PSG sample.  Low, medium and  high density environments are 
described by $\rho_{\rm{g}} < 0.21$, $ 0.21 <\rho_{\rm{g}} < 0.58$  and $\rho_{\rm{g}} > 0.58$ respectively
\citep{schawinski07a}.
The low density class of environment can be likened to the galaxy field environment.  Similarly,  the 
medium and high density classes can be compared to the group (or cluster outskirts) and cluster 
environments, respectively.

\begin{table}
\begin{center}
\caption{Fractions of galaxies in three different environment classes for different types of galaxies.  }
\label{envtab}
\footnotesize{
\begin{tabular}{lccc}
\hline
\hline
 Galaxy type & $\rho_{\rm{g}} < 0.21$ & $ 0.21 <\rho_{\rm{g}} < 0.58$  &$\rho_{\rm{g}} > 0.58$\\
\hline
Early-type & 45\% (4034) & 29\% (2561) & 26\% (2333)\\
Intermediate & 53\% (11866) & 27\% (6022) & 20\% (4592)\\
Late-type  & 56\% (9124) & 28\% (4555) & 16\% (2566)\\
PSG & 50\% (40) & 26\% (21) & 24\% (19) \\
\hline
\hline
\end{tabular}}
\end{center}
\end{table}

We find that 50\% of our sample reside in the low density environment, while 26\% and 24\% reside in 
the medium and high density environments, respectively.   In high density environments, the fraction of 
PSGs is similar to that of early-types, while in low density environments, the fraction of PSGs is in 
between that of intermediate- and early-type galaxies. As can be seen from Table~\ref{envtab}, 
the number fractions of early- and late-type galaxies in the low and high density environments differ by 
$\sim$10\% whereby a greater fraction of early-types are found in high density environments and a greater 
fraction of late-types are found in low density environments.  This greater fraction of early-type
galaxies at higher densities recapitulates the well-known morphology--density relation found by
\citet{dressler80} and more recent studies such as \citet{bamford09}. Similar to the PSG sample, the 
intermediate-type galaxies appear to have fractions which are in between those found for early- and late-types. 
In addition, the fractions of galaxies residing in medium density environments are roughly the same 
regardless of galaxy types.
It should be noted that KS tests comparing the $\rho_{\rm{g}}$ distributions of different galaxy types 
confirm these results.  Hence, we find that the PSG distributions across all three environment 
classes are similar to those of the early- and intermediate-type galaxies. This is consistent with the 
fact that $\sim90$\% of the PSG sample consists of galaxies classified as early- or intermediate-type 
galaxies.  Our results are also consistent with those of \citet{blake04} who found that the local 
environments of E+A galaxies follow that of the general galaxy population.

\section{Discussion}

Our results show that local post-starburst galaxies represent one population of galaxies which 
currently occupies a well-defined position in the low-stellar-mass end of the ``green valley''
and is rapidly transitioning onto the low-mass end of the ``red sequence'' unless star formation
resumes within the  transitioning period of $\sim1$ Gyr.  The duration within which the individual 
PSG spends in the ``green valley'' is probably only on the order of $\sim1$ Gyr (the timescale for 
the fading of the Lyman continuum from the current generation of young stars) because structurally,
 the stellar concentration of the PSGs within the ``green valley'' already closely resemble those
 of low-mass early-type galaxies even though star formation has only been truncated recently. 
Our transition timescale concurs with recent findings of transition times on the order of $\sim1$ 
Gyr between the ``blue cloud'' and the ``red sequence'' \citep[e.g.\ ][]{kaviraj11,kaviraj07, schawinski07b}.
These results are  consistent with the idea of downsizing in the sense that larger objects have mostly
been formed at earlier epochs, while smaller objects are still being formed at later epochs.
Our results are therefore comparable to those of \citet{wild09} who found the mass density of PSGs to be
 230 times lower at $z\sim 0.07$ than at $z\sim 0.7$.


Current galaxy evolution models often suggest that feedback from an active galactic nuclei (AGN) could 
provide the means to quench and truncate the star formation history of a massive galaxy 
\citep[e.g. ][]{silk98, kaviraj05,croton06}.
Incidentally, our observation of a stellar mass limit in our PSG sample coincides with the findings of 
\citet{schawinski10a} who found that the AGN duty cycle peaks at the low mass-end of the ``green valley'' 
and that the low-mass early-type AGN hosts appear to have post-starburst properties.  However, apart from two 
PSGs which exhibit spectral properties of LINERs, we do not observe any spectral signatures of AGN within 
our PSG sample.

The connection between AGN and merger/interactions have been discussed in the context of mergers 
inducing in-flows of gas that fuel star formation and the black hole \citep[in the central regions; e.g.\ ][]{canalizo06},
 while feedback from the AGN is predicted to quench star formation by re-heating the cold gas and expelling 
much of it in AGN-driven winds \citep[e.g.\ ][]{dimatteo05,thacker06}.  \citet{tremonti07} found evidence for 
these winds in 10 out of 14 PSGs at $z=0.6$.  They hypothesised that the observed gas outflows in these 
galaxies suggest that AGN feedback may play a role in quenching star formation in PSGs.
Compared to the PSG sample of \citet{tremonti07}, our local PSGs are redder and not as massive.
Therefore, following the results of \citet{yan09} and \citet{wild09}, we hypothesise that the evolution of our 
local PSG sample is likely to be different to that of PSGs at higher redshifts. Consistent with the results of 
\citet{kaviraj07}, it is unlikely that AGN feedback will be a dominant quenching process for star formation
in local PSGs.


\section{Summary}
In this paper, we have presented a study of local post-starburst galaxies (PSGs) using the photometric and spectroscopic
observations from SDSS in conjunction with the results from the Galaxy Zoo 1 project. We find that:

\begin{itemize}
\item The local population of PSGs occupy a well-defined space on the colour--stellar mass diagram, most notably in the
low-mass end of the green valley below the transition mass \citep[log $M_{\star} <$ 10.5 M$_{\odot}$; ][]{kauffmann03}
 thought to be the mass division between low-mass star-forming galaxies and high-mass passively-evolving bulge-dominated
galaxies.  Consistent with the idea of galaxy formation downsizing where smaller galaxies form at later epochs, we think 
that the local PSGs  will contribute to the build-up of the low-mass end of the red sequence if star formation has indeed
ceased in these galaxies.

\item Consistent with previous studies \citep[e.g. ][]{blake04}, we find that local environment of local PSGs follow that 
of the general galaxy population within the same volume. 

\item Using the morphological classifications from Galaxy Zoo, we are able to study the distributions
of morphologies in our PSG sample in comparison to those of 47,573 galaxies in our full galaxy sample within the same
 local volume. Although a majority of our local PSG sample appears to have intermediate-type morphologies which
are neither early- nor late-type morphologies, we find that  the stellar structural morphology (as described by {\tt{fracDev}}) 
of the local ``green valley'' PSGs to be very similar to that of low-mass early-type galaxies in the ``red sequence'' even 
though star formation has only recently ceased.  Therefore, unless star formation resumes, we hypothesize that the 
local PSGs  will evolve out of the ``green valley'' in  $\sim1$ Gyr  onto the ``red sequence'' as soon as the young stellar 
population from the most recent episode of star formation fades.



\end{itemize}

\vspace{1cm}

\chapter{\bf{Acknowledgments.}}
Galaxy Zoo acknowledges support from The Leverhulme Trust.
OIW is the recipient of a Super Science Fellowship from the Australian
Research Council.  Support for the work of KS was provided by NASA through 
Einstein Postdoctoral Fellowship grant number PF9-00069 issued by the Chandra 
X-ray Observatory Center, which is operated by the Smithsonian Astrophysical 
Observatory for and on behalf of NASA under contract NAS8-03060. 
KLM acknowledges funding from the Peter and Patricia Gruber Foundation
as the 2008 Peter and Patricia Gruber Foundation International
Astronomical Union Fellow, from a 2010 Leverhulme Trust Early Career
Fellowship and from the University of Portsmouth and SEPnet
(www.sepnet.ac.uk). The authors also thank the anonymous referee
 for the constructive comments which has improved this paper.

\bibliographystyle{mn2e} 
\bibliography{mn-jour,paperef}

\end{document}